# HapTable: An Interactive Tabletop Providing Online Haptic Feedback for Touch Gestures

Senem Ezgi Emgin, Amirreza Aghakhani, T. Metin Sezgin, Cagatay Basdogan

**Abstract**—We present HapTable; a multi-modal interactive tabletop that allows users to interact with digital images and objects through natural touch gestures, and receive visual and haptic feedback accordingly. In our system, hand pose is registered by an infrared camera and hand gestures are classified using a Support Vector Machine (SVM) classifier. To display a rich set of haptic effects for both static and dynamic gestures, we integrated electromechanical and electrostatic actuation techniques effectively on tabletop surface of HapTable, which is a surface capacitive touch screen. We attached four piezo patches to the edges of tabletop to display vibrotactile feedback for static gestures. For this purpose, the vibration response of the touch screen, in the form of frequency response functions (FRFs), was obtained by a laser Doppler vibrometer for 84 grid points on its surface. Using these FRFs, we have developed a new technique to display localized vibrotactile feedback on the surface for static gestures. For dynamic gestures, we utilize electrostatic actuation technique to modulate the frictional forces between user's fingers and tabletop surface by applying voltage to the conductive layer of the touch screen. To our knowledge, this hybrid haptic technology is one of a kind and has not been implemented or tested on a tabletop. It opens up new avenues for gesture-based haptic interaction not only on tabletop surfaces but also on touch surfaces used in mobile devices with potential applications in data visualization, user interfaces, games, entertainment, and education. Here, we present two examples of such applications, one for static and one for dynamic gesture, along with detailed user studies. In the first one, user detects the direction of a flow, such as that of wind or water, by putting her/his hand on the surface and feels a vibrotactile stimulus traveling underneath it. In the second example, user rotates a virtual knob on the surface to select an item from a menu while feeling the knob's detents and resistance to rotation in the form of frictional haptic feedback.

✦

## 1 INTRODUCTION

In contrast to personal computers utilizing indirect input devices such as mouse and keyboard, interactive tabletops allow users to directly manipulate digital content via touch gestures. They intuitively couple gesture input with direct graphical output, which requires minimal learning and enables natural interaction. They also provide a large horizontal surface, allowing multiple users to collaborate simultaneously and interact with each other [1]. However, they lack the physicality of an interaction as experienced with the input devices, and consequently, require full visual attention of the user, which is tiring and results in deterioration in task performance [2].

One of the key senses for interaction is haptics. Haptic feedback is known to improve task performance (in terms of completion time and precision) and realism. It also helps to reduce cognitive load and enables representation and digestion of complex data more easily [3]. There is an ongoing effort in research community for adding haptic feedback to interactive tabletops and surface displays. One such effort is to develop shape-changing surfaces. For example, FEELEX is made of an array of 36 linear actuators, each moving individually in vertical direction to project the surface contour of a digital image on a flexible surface [4]. Lumen is an array of movable light guides whose height and color can be controlled individually to create images, shapes and physical motions [5]. The motion of each light guide is controlled by a string, made of shape memory alloy (SMA), attached to the guide. More recently, Follmer et al. [6] developed inFORM, which enables dynamic affordances, constraints, and actuation of passive objects. This system utilizes 900 motorized pins (30x30) to actuate 150 boards moving up and down to render dynamic shapes on the surface. Haptic feedback is displayed to the user by adjusting the stiffnesses of pins via a PID control. As stated by the authors, shape changing displays are currently not practical due to their large size and cost of manufacturing.

Another common line of effort for displaying haptic feedback through a touch surface is to utilize electromechanical or electrostatic actuation. Poupyrev et al. attached four piezo actuators to the corners of a pen-based touch display, in between LCD and protective glass panel, to convey vibrotactile haptic feedback to users by varying amplitude and frequency of input signal [7]. The results of their user study showed that subjects preferred haptic feedback when it was combined with an active gesture, such as while dragging a slider or highlighting a text using the pen interface. Jansen et al. developed Mud-Pad, a device that utilizes magnetorheological fluid combined with small electromagnets placed under the display surface [8]. The fluid's physical properties are altered using electromagnets, thus the frictional properties of the

---

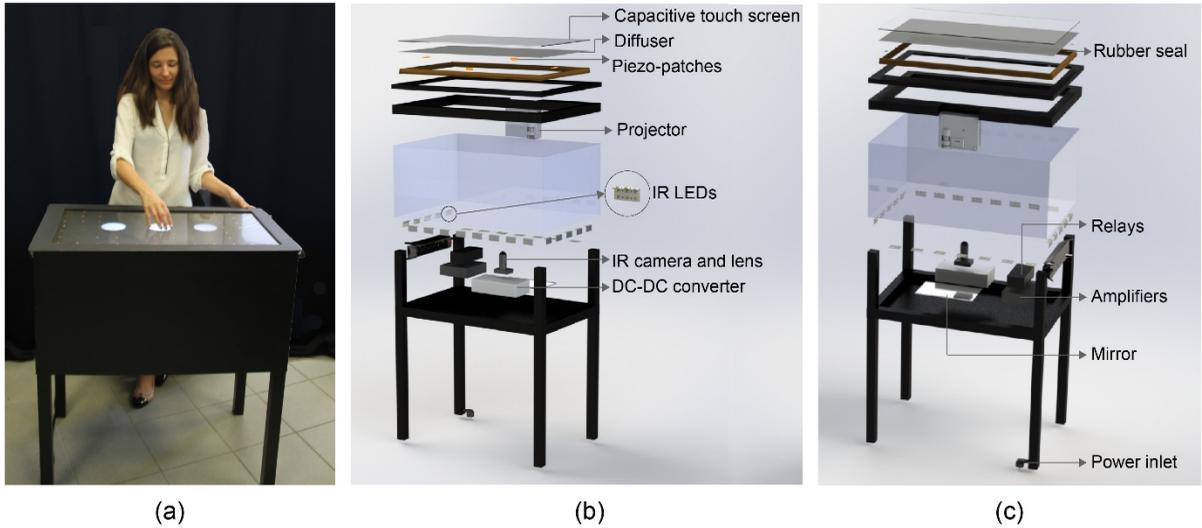

Fig. 1. (a) User performing a gesture on the proposed table to interact with a digital scene while receiving suitable haptic feedback. The hardware components of the proposed table are shown in (b) rear and (c) front views.

surface are controlled to provide active tactile feedback to user. Due to the electromagnets under the surface, this system is not compact and requires visual projection from top. Bau et al. presented TeslaTouch, a capacitive touchscreen utilizing electrostatic actuation [9]. The device controls the attractive electrostatic force between user finger and display surface by modulating the voltage applied to the conductive layer of the screen. Yamamoto et al. also used the same principle in a tactile telepresentation system to realize explorations of remote surface textures with real-time tactile feedback to user [10].

To display a rich set of haptic effects for the gestures performed on a tabletop, we integrated electromechanical and electrostatic actuation techniques effectively on HapTable (Fig. 1). We attached four piezo patches to the edges of the table's interaction surface to control its out-of-plane vibrations and display localized vibrotactile haptic feedback to user for static gestures. For tabletop interactions using dynamic gestures, we convey haptic feedback to user via electrostatic actuation technique introduced in [9, 10]. We modulate the frictional forces between user's finger(s) and our tabletop surface (a large-size surface capacitive touch screen, also referred to as touch screen in the text) in real time according to the dynamic gesture performed on the surface.

Using this hybrid actuation approach, the type of haptic feedback that can be displayed through HapTable varies in complexity from simple frictional effects to more complex localized vibrotactile flow effects. Our particular approach for creating localized vibrotactile effects on HapTable requires vibrational characterization of its touch surface and intense precomputations. To demonstrate how haptic feedback can improve user's interactions with HapTable, we present two example applications, supported by detailed user studies. In the first example, user detects the direction of a travelling vibrotactile flow, mimicking a flow of wind or water, by placing her/his hand on the surface. In the second example, user rotates a virtual haptic knob using two fingers to select an item from a menu while feeling the detents of the knob and receiving frictional feedback according to her/his rotational movement.

This paper is organized as follows: Section 2 introduces our table and its hardware components. Section 3 introduces our methods for recognizing static and dynamic hand gestures in real time. Section 4 discusses our haptic rendering methods utilizing electromechanical and electrostatic actuation techniques. Section 5 presents our user studies, investigating how haptic feedback may augment tabletop interactions triggered by static and dynamic gestures. The results of the user studies are discussed in Section 6. The final section concludes this paper and elaborates on our future work.

## 2 DESIGN OF HAPTABLE

HapTable system consists of three main modules: gesture detection, visual display, and haptic feedback (Fig. 1).

Gesture detection module is responsible for registering and detecting high resolution images of static and dynamic hand gestures performed on HapTable surface. Although there are touch surfaces commercially available in the market for detecting finger and/or hand gestures, they may potentially interfere with our haptic feedback module and may not capture hand contour in sufficient detail for correct recognition of hand gestures. For example, the piezo actuators that are used to generate vibrotactile haptic effects in HapTable may interfere with the travelling sound waves utilized in surface acoustic touch sensors to detect finger poisiton. Similarly, infrared touch frames have occlusion problems and are not good at detecting hand contour. For these reasons, HapTable uses rear diffused illumination (Rear DI) to register hand poses [1]. The tabletop surface is evenly illuminated with wide-angle infrared LEDs (50-Module IR Kit, Environmental Lights) in configuration of three rows: top row is placed parallel to the table surface to illuminate its edges, whereas the remaining two rows are perpendicular to the first row to illuminate the surface center (Fig. 1b). When a user touches the HapTable surface, light is reflected from contact points and captured by an infrared camera (Eye 3, PlayStation). This camera captures 60 frames per second with a resolution of 640 x 480 pixels.

The visual display of digital images on the tabletop surface is achieved by a projector (B1M, Asus). We selected this projector specifically because it does not emit infrared light that may interfere with gesture detection module, and it has a short throw distance that allows minimal table depth. The throw distance is extended by using an additional mirror (Fig. 1c), allowing users to interact with HapTable even in a sitting position.

The haptic feedback module integrates electromechanical and electrostatic actuation techniques to display a wide range of haptic effects while users interact with digital images and objects through static and dynamic gestures. These two actuation techniques complement each other. For static gestures, HapTable displays vibrotactile haptic feedback to user. Four piezoelectric patches (PI-876.A 12, Physik Instrumente, 61 x 35 x 0.5mm) were attached beneath the touch surface to generate mechanical vibrations on the surface. The propagation of the vibrations from the touch surface to the table itself is prevented by rubber seals placed under the tabletop surface (Fig. 1c). We utilize electrostatic haptic feedback for dynamic gestures. This technology does not use any form of mechanical actuation but tactile sensations can be created by controlling frictional forces between the tabletop surface and user's fingers. In order to generate friction on the surface based on electrostatic actuation, a large surface capacitive screen (SCT-3250, 3M, 743.46 x 447.29 x 3.18mm) is used as the touch surface of the table. When a periodic voltage is applied to the conductive layer of the screen, normally used for sensing finger position, an attractive electrostatic force develops between finger skin and the screen surface in the normal direction. This electrostatic force is small and cannot be sensed directly by finger while it is stationary on the surface. However, if finger slides on the surface of the touch screen, a resistive frictional force is felt by user in tangential direction.

To control the voltage transmitted to the piezo patches for vibrotactile haptic feedback and also to the touch screen for electrovibration independently; a sound card, two high-voltage amplifiers (E413.D2, Physik Instrumente, Gain: 50), and two solid state relay arrays (Yocto-MaxiCoupler, Yoctopuce) are used (Fig. 2). The haptic signals generated by the left and right output channels of

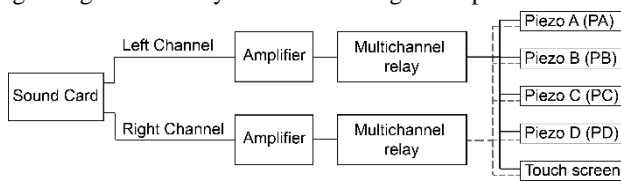

Fig. 2. Schematic showing how voltage is transmitted to piezo patches for vibrotactile feedback, and to touch screen for electrostatic feedback. This design allows HapTable to send different stimulus to an individual or combination of piezo patches and to the electrostatic touch screen.

the sound card of a personal computer (PC) are first transmitted to the high-voltage amplifiers. Each amplifier's positive output is connected to a multichannel solid-state relay array, controlled and powered by the USB ports of PC. This relay is fast, can switch voltages up to 350 Vpp for small loads of current (up to 100 mA), and does not require any external power. The outputs of these relays are connected to each piezo patch and the touch screen, as shown in Fig. 2. This architecture enables us to excite any number of patches and the touch screen simultaneously. However, we can only apply independent voltage signals to at most one piezo patch and the touch

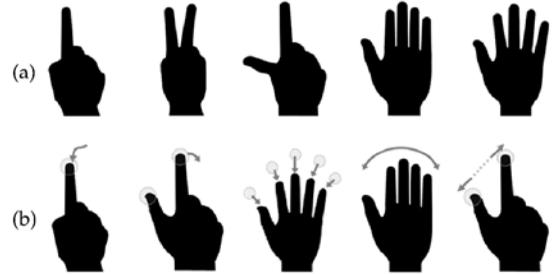

Fig. 3. Selected gestures for HapTable: (a) static (from left to right: 1-finger, 2-finger, L-shape, hand with closed and open fingers), and (b) dynamic (from left to right: dragging, rotation, spread/pile, wipe, zoom in/out) gestures.

screen since the sound card has only two output channels.

## 3 GESTURE RECOGNITION

An important feature of HapTable is the real-time recognition of hand gestures performed on its touch surface. In general, hand gestures performed on touch surfaces can be classified as: (a) static and (b) dynamic. In static gestures (e.g. pressing a button, pointing an object), the hand has fixed position and orientation, while in dynamic gestures (e.g. dragging a folder, rotating a virtual knob), it has time-varying position and orientation. In [11], authors differentiate static gestures from dynamic ones by examining the positional change of hand pose for a specific time window. If it does not change in time, then the gesture is classified as static. However, this algorithm recognizes the trajectory of hand gesture, rather than hand contour. Authors in [12] examine the hand shape and recognize a gesture in 1.5 seconds using Self-Growing and Self-Organized Neural Gas (SGONG) algorithm. However, this duration is long and not feasible for real-time haptic interactions on our table. According to [13], users expect a response in less than 20 milliseconds in interactive systems. Hence, to provide real-time haptic feedback for a gesture, our system has to recognize the gesture at its early stage of evolution. For this reason, it is crucial for us to select simple but discriminative features for gesture recognition. Since the user can make gestures anywhere on the table, the selected features should also be independent of translation, orientation, and hand size.

We developed a simple yet efficient algorithm that can recognize pre-selected five static (Fig. 3a) and five dynamic hand gestures (Fig. 3b). The first step in our algo-rithm is to decide if a gesture is static or dynamic based on the change in position and orientation of the hand. Then, the camera images are sent to the relevant classifier accordingly for further processing.

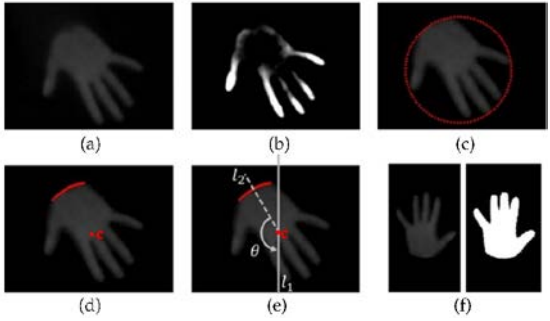

Fig. 4. The steps of rotating a hand pose according to a reference edge: (a) image frame obtained by the infrared camera is subtracted from the background, (b) a high-pass filter is applied to reveal parts contacting the table surface, (c) smallest circle enclosing the hand posture is found, (d) the wrist (red arc) and the center point of the hand (point c) are determined, (e) the angle ($\theta$) that the wrist makes with the reference line ($l_1$) is estimated, and (f) the adjusted hand pose perpendicular to the reference edge is obtained.

If a hand gesture is static, it is rotated into a canonical orientation with respect to a reference edge and wrist is removed from the pose since users can approach the table from any side (Fig. 4). Aligning a hand pose with respect to the reference edge is achieved as follows: we first apply a high-pass filter to the raw image in order to highlight the parts that are in contact with the table (Fig. 4b). Then, the smallest circle enclosing the highlighted hand (i.e. bounding circle) is calculated (Fig. 4c). The arc intersecting the bounding circle is defined as the wrist (Fig. 4d). To rotate the hand and make it perpendicular to the reference edge, the angle $\theta$ between this reference edge ($l_1$ in Fig. 4e) and the line connecting the midpoint of the wrist to the center of the bounding circle ($l_2$ in Fig. 4e) is calculated. If the hand pose is rotated by an angle of $\theta$ degrees in counter-clockwise direction about the center point (c in Fig. 4e), it becomes perpendicular to the reference edge as shown in Fig. 4f. Then, the silhouette of the hand posture is recognized via Fourier descriptors [14] using Support Vector Machine [15].

Compared to a static hand gesture, we need to recognize a dynamic hand gesture at its early stage of evolution to provide the user with haptic feedback immediately. For this reason, our algorithm utilizes the first four frames of a dynamic gesture for feature extraction. In addition to Fourier descriptors, the number of fingers in contact and the trajectory of contact points are also used as discriminating features to classify the dynamic gestures.

The gesture evaluation experiments were conducted with 5 subjects (2 males and 3 females). A sketch for each gesture was presented to the subjects, and they were asked to repeat this gesture 40 times in different positions and orientations on touch surface. Hence, the total number of gestures performed on the tabletop surface was 2000 (5 subjects x 10 gestures x 40 repetitions). We trained and tested our recognition algorithm using two-fold cross validation approach. Recognition rates of 98% and 91% were achieved for static and dynamic gestures, respec-tively, without compromising the responsiveness of the system.

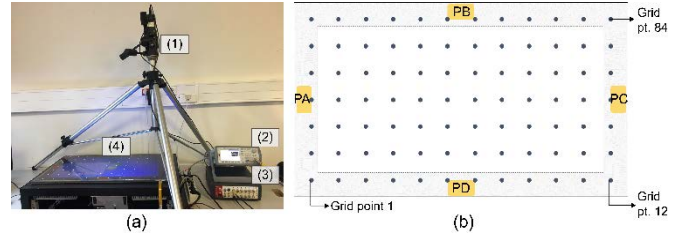

Fig. 5. (a) The setup for vibration measurements: (1) Laser doppler vibrometer (LDV), (2) signal generator, (3) signal analyzer, and (4) touch surface. (b) The tabletop surface was divided into 84 equally-spaced grid points for the measurements (7 rows by 12 columns). PA, PB, PC, and PD represent the piezo patches glued to the edges underneath the touch surface.

## 4 HAPTIC FEEDBACK

In our table, haptic feedback is displayed to user according to the type of gesture she/he performs and the digital content she/he is interacting with.

### 4.1 Haptic Feedback for Static Gestures

We display localized vibrotactile haptic feedback for static gestures. For this purpose, we first construct a vibration map of the touch screen in advance and then display haptic feedback for the gesture accordingly during real-time interaction. To construct the vibration map of touch screen, we divided its surface into 84 grid points (7 rows by 12 columns, Fig. 5b). The size of each grid was 6 x 6 cm. The out-of-plane vibrations at each grid point were measured when each piezo patch was excited individually and when all patches were excited together. For this purpose, a linear sine sweep signal, varying in frequency from 0 to 625 Hz, was generated by a signal generator, amplified by one of the high-voltage amplifiers in our setup (E413.D2, Physik Instrumente, Gain: 50), and then transmitted to the terminals of the piezo patches. A Laser Doppler Vibrometer (LDV, PDV-100, Polytec) was used to measure the out-of-plane vibrations at each grid point (Fig. 5a). A signal analyzer (NetDB, 01dB-Metravib) was used to record and analyze the signals coming from LDV and the signal generator. Having defined the signal generator's output as the reference channel in the signal ana-lyzer, the experimental frequency response functions (FRFs) between the velocity output and piezoelectric voltage input were obtained. The same process was repeated 3 times for the cases when each piezo patch was active and when all piezo patches were active together in parallel configuration. The velocity FRF of each grid point was estimated by averaging the data of three full sweeps, and then converted to displacement FRFs (Fig. 6). The averaged FRFs for patches PA and PC, and PB and PD are similar due to their symmetrical configurations on HapTable surface, as shown in Fig. 5b.

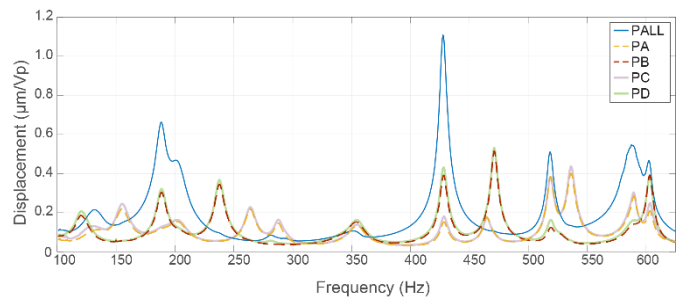

Fig. 6. Displacement FRFs for five excitation cases: patches PA, PB, PC, and PD are excited individually and together in parallel configuration (PALL).

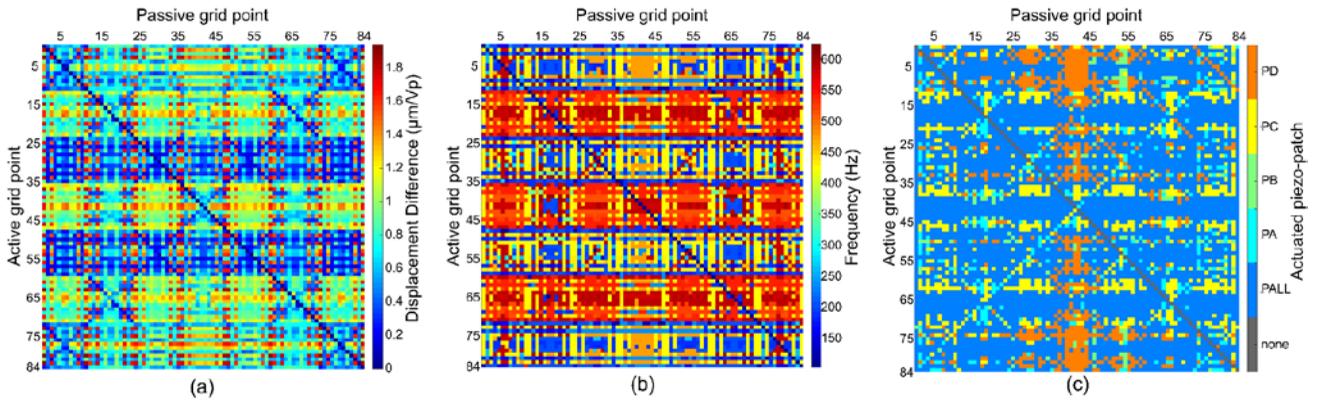

Fig. 8. Excitation lookup tables for HapTable: To create a vibrotactile flow from an "active" grid point to a "passive" one, our haptic rendering algorithm acquires maximum displacement difference in vibration amplitudes from table (a), the corresponding excitation frequency from table (b), and the actuator ID (i.e. which actuator to use to create that displacement difference at the corresponding excitation frequency) from table (c).

In order to display localized vibrotactile haptic feedback on the touch surface, we utilize the five FRF functions (one for each piezo patch and one for all together) of 84 grid points (referred to as "vibration map" in the text). For each grid point, there is an excitation frequency at which the amplitude is maximum. If the surface is excited at this frequency, a localized haptic effect can be generat-ed at and around that point. Furthermore, it is even possible to generate directional vibrotactile flow between any two grid points on the screen by simply switching between the excitation frequencies corresponding to the maximum difference in their displacements. For non-grid points, FRFs can be estimated by bilinear interpolation.

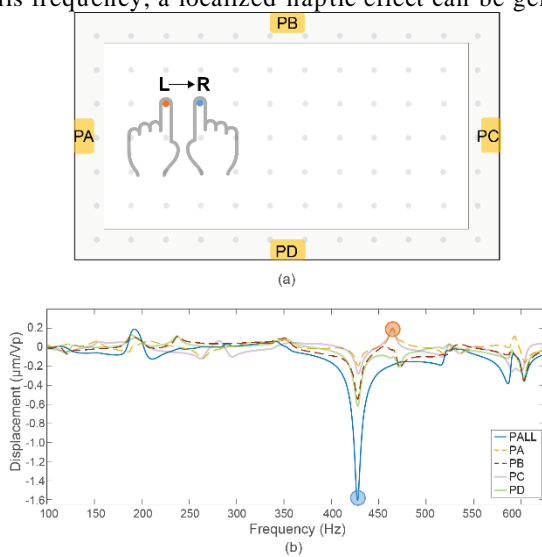

Fig. 7. To create a flow from point L to R on the touch screen (a), the FRF graph of point R is subtracted from that of point L for the five excitation cases: when piezo-patches are excited individually and all together (b). The maximum and minimum differences in vibration amplitude are marked by orange and blue circles, respectively.

For instance, consider the two points illustrated in Fig. 7a, contacted by the index fingers of both left and right hands. To generate a vibrotactile flow from point L to point R, two excitation frequencies are carefully chosen from the FRF plots (Fig. 7b) so that point L is the active point and has higher vibration amplitude than that of point R in the first part of the stimulus, and vice versa in the second part of the stimulus. The difference in FRF plots of points L and R are shown in Fig. 7b. This plot shows that the vibration amplitude of point L makes a maximum difference with that of point R (marked with orange circle) at 465 Hz when the piezo patch PA is active. Similarly, the blue circle indicates that the vibration amplitude of point R is significantly larger than that of point L at 428 Hz when all piezo patches are excited simultaneously (PALL). Hence, PA is actuated first and then PALL. Fast solid-state relays shown in Fig. 2 are used to switch between the actuators PA and PALL during the display of haptic stimulus.

In order to extend the vibrotactile flow concept to all grid points on the surface efficiently (and hence to all points on the surface through bilinear interpolation), we have developed a sophisticated preprocessing approach. We construct and store three lookup tables (Fig. 8), which are used to determine the excitation parameters during user interaction in real time. These tables store the maximum difference in vibration amplitudes of the grid points (Fig. 8a), the corresponding excitation frequencies (Fig. 8b), and the actuator ID (Fig. 8c; either the patch PA, PB, PC, PD, or PALL). For example, if the points L and R, shown in Fig. 7a, are selected as active and passive points respectively (they correspond to the grid points 51 and 52 on the surface) and inputted to the tables, a maximum vibration difference of 0.201 μm/Vp (Fig. 8a) at 465 Hz (Fig. 8b) is returned. This difference in amplitude is obtained when the surface is actuated by piezo patch PA (Fig. 8c). On the other hand, if point R is active and point L is passive, the vibration difference now becomes 1.607 μm/ Vp (Fig. 8a), which is obtained at a different actuation frequency of 428 Hz (Fig. 8b) and when the surface is actuated by PALL (Fig. 8c). Using this information, a haptic stimulus is created for displaying a vibrotactile flow effect from left to right as shown in Fig. 9b. First part of the stimuli excites PA to create a substantially high vibration displacement at point L compared to that of at point R, whereas the second part excites PALL to accomplish vice versa. The amplitudes on both parts of the stimulus are adjusted according to the human sensitivity to vibrotactile excitation [16, 17] to create an equivalent haptic effect in magnitude (Fig. 9b). Then, a linear amplitude modulating envelope is applied to the beginning and end of the signals in each part of the stimulus to make the transitions smoother during the activation and deactivation periods of the piezo pathces (Fig. 9b and 9c). We demonstrate in Section 5 how this technique can be used to create a directional vibrotactile flow between the index fingers of left and right hands and also underneath one hand placed on the tabletop surface.

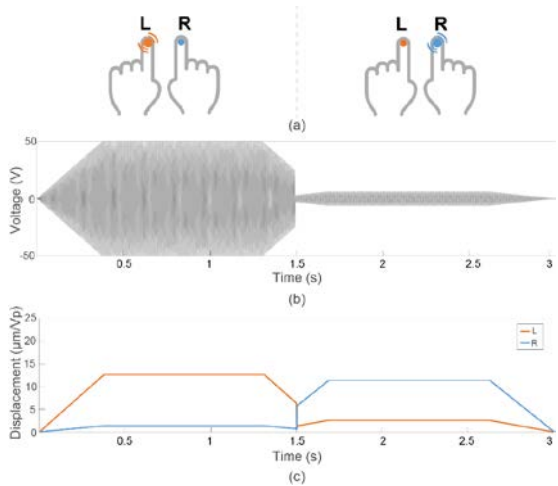

Fig. 9. (a) To create a vibrotactile flow from left (L) to right (R), we first actuate the surface at a certain frequency such that the vibration amplitude of point L is significantly higher than that of the point R, and then actuate the surface at a different frequency such that the vibration amplitude of point R is significantly higher than that of the point L. Note that different piezo actuators could be used to play the voltage signals shown on the first (0 – 1.5 seconds) and second (1.5 – 3.0 seconds) parts in the figure.

### 4.2 Haptic Feedback for Dynamic Gestures

To display haptic feedback for dynamic gestures, we modulate the frictional forces between user's fingers and the touch screen used as the tabletop surface of HapTable. When an alternating current voltage is applied to the conductive layer of a touch screen, electrostatic attraction force ($f_e$) develops between fingers and the surface of touch screen (Eq. 1). The magnitude of this attractive force is governed by the applied voltage, V(t), contact area, A, permittivity of the vacuum, insulating layer of the touch screen, and outer finger skin ($\varepsilon_0$, $\varepsilon_i$, and $\varepsilon_s$ respectively), and the insulator and outer skin thicknesses ($t_i$, $t_s$), as written in below [18]:

$$f_e = \frac{\varepsilon_0 V^2(t) A}{2(t_i + t_s)(\frac{t_i}{\varepsilon_i} + \frac{t_s}{\varepsilon_s})} \quad (1)$$

The magnitude of this attractive force is too small to be perceived by a stationary finger on the touch surface. However, it results in a perceivable change in frictional force in tangential direction when human finger moves on the surface.

$$\vec{F}_f = \mu(\vec{F}_N + \vec{f}_e) \quad (2)$$

By controlling the frequency, amplitude, and the waveform of the applied voltage, it is possible to create different haptic effects on the surface [9]. In the upcoming section, we demonstrate the use of this technology in a case study involving a virtual knob, which is rotated by two fingers on the surface.

## 5 USER STUDIES

We demonstrate the functionality of HapTable via two example applications supported by detailed user studies. In the first one, as an exemplar for static gestures, we render localized directional vibrotactile flow between index fingers of two hands and also underneath a hand placed on the surface. We investigate if users can differentiate the direction of vibrotactile flow. As an example for dynamic gestures, we haptically render a virtual knob on the surface. The user receives frictional haptic feedback as she/ he rotates the knob using two fingers in order to select an item from a pull-down menu. We investigate if haptic feedback improves task performance and the user's subjective sense of performing the task successfully.

### 5.1 Vibrotactile Flow

The vibrotactile flow, i.e. haptic illusion of apparent tactile movement on human skin, was investigated by Sherrick and Rogers in 1960s [19]. They attached two vibration motors on user's thigh separated by a distance and then adjusted the stimulus duration and the delay between the actuation times to create an effect of a traveling haptic stimulus. They showed that stimuli duration and the interstimulus offset interval (ISOI, i.e. the temporal interval between the offset of one vibration to the onset of another one), are the key parameters that affect the subjects' haptic perception. Tan and Pentland [20] and Israr and Poupyrev [21] extended this concept to 2D surfaces by placing an array of vibration motors on the cushion of a chair to create directional tactile strokes. In a separate study, Israr and Poupyrev [22] investigated the control parameter space for producing reliable continuous moving patterns on forearm and back. The results of their user study showed that ISOI space for the forearm was influenced by both the motion direction and spacing of the actuators, whereas ISOI space for the back was affected only by the direction of actuation. Arasan et al. [23] applied the apparent tactile motion to a pen-based stylus that can be used with a tablet or a mobile device. Two vibration motors were placed at the proximal and distal ends of the stylus to create a tactile illusion of traveling wave along its long axis (up to down or down to up). They demonstrated the potential applications of this stylus in computer games and data visualization.

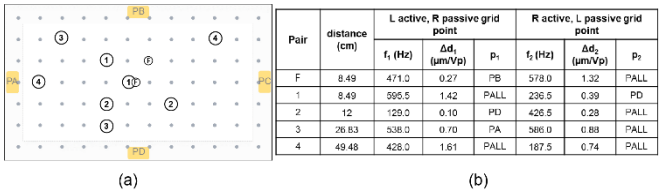

(a)                           (b)

Fig. 10. (a) The locations of the test pairs selected for the familiarization (pair: F-F) and the actual experiment (pairs: 1-1, 2-2, 3-3, 4-4) sessions. (b) Distance between the test points for each pair, actuation frequencies ($f_1$, $f_2$), the piezo patches used for actuation ($p_1$, $p_2$), and the displacement differences between the test points ($\Delta d_1$, $\Delta d_2$) are given in the table on the right.

The illusion of tactile apparent motion can also be generated by amplitude or frequency modulation. Kim et al. [24] created the sensation of a traveling wave between two vibration actuators embedded in a cell phone by adjusting the magnitude and timing of the actuators. Lim et al. [25] used frequency modulation to create a vibrotactile flow between two hands holding a tablet equipped with vibration motors. Kang et al. [26] used piezo-patches glued to the short edges of a tablet-size glass plate for creating a vibrotactile flow between them via frequency modulation. They modulated the frequency from zero to the first mode of the plate in order to create an illusion of moving tactile stimulus from one short edge of the plate to the other on the opposite side.

## 5.2 Vibrotactile Flow Experiments

### 5.2.1 Experiment 1: Vibrotactile flow between two points

We conducted an experiment with 5 subjects (2 female, 3 male) having an average age of 31.4 years (SD = 6.9) to investigate if they can detect a directional vibrotactile flow between two grid points on the touch surface. In order to create a directional vibrotactile flow, we used an amplitude-modulated voltage signal having two parts (see the profile in Fig. 9). Each part was played by the appropriate piezo actuator. The frequency of the signal in each part and the actuator that plays the signal were carefully selected from the excitation lookup tables, as discussed in the previous section. Subjects placed their left and right index fingers on the designated locations (test points) of the tabletop surface (Fig. 10a) and were asked the perceived direction of vibrotactile flow: left to right hand, or right to left hand. All subjects were asked to wear active noise-canceling headphones playing white noise to prevent them hearing auditory cues caused by the vibrations. Only one subject was left-handed. Prior to the experimentation, all subjects were informed about the nature of the experimental procedure.

Experiment started with a familiarization session, consisting of 10 trials (5 repetitions for each direction in random order) performed with a single pair (pair F-F in Fig. 10a). Subjects were allowed to replay each stimulus as many times as needed during the familiarization session. Afterwards, the actual experiment was conducted with 4 pairs of test points, located at different regions on the touch screen. Subjects could replay each stimulus only once in the actual experiment. More information about the selected test pairs, the physical distance between them, the actuators played the voltage signal in each part of stimulus, and the difference between their vibration amplitudes are reported in Fig. 10b. All vibration amplitudes in the experiment were above the absolute vibrotactile threshold of human finger for the excitation frequencies listed in Fig. 10b. The experiment consisted of 40 trials (4 pairs x 2 directions x 5 repetitions) displayed in random order.

All subjects identified the direction of vibrotactile flow with a perfect accuracy of 100% for all pairs. The results of this experiment showed that subjects could easily differentiate the direction of vibrotactile flow with their index fingers even if the test points are diagonal to each other, as in the pairs of 1-1, 3-3, and 4-4.

### 5.2.2 Vibrotactile flow under hand

Based on the encouraging results of the first experiment, we expanded our study to investigate directional vibrotactile flow under a hand placed on the touch surface. We assumed that human hand covers an area of 12 x 12 cm. We divided this area into nine equal squares (a), and each square is further divided into 15 x 15 subgrid points for finer resolution (FRFs for these points were calculated in advance using bilinear interpolation). Similar to the concept of active and passive points introduced in the first experiment, active and passive squares are defined in this experiment. A square is assumed to be active if at least half of its subgrid points have a sufficiently high vibration (three JND above the human vibrotactile threshold for the applied excitation frequency). The threshold and just noticeable difference values for different excitation frequencies were obtained from the human vibrotactile sensitivity curve reported in [16, 17].

Using these active and passive squares, we aimed to generate a vibrotactile flow in horizontal and vertical directions. As in the case of the first experiment, this required to select proper excitation frequencies from the FRFs. In order to create a directional vibrotactile flow, the active and passive squares should be symmetric with respect to the horizontal (vertical) axis passing through the center point of the area representing hand (point C in Fig. 11a) for vertical (horizontal) flow. Fig. 11 illustrates example vibration maps that are acceptable (Fig. 11b, d) and unacceptable (Fig. 11c, e) based on our algorithm.

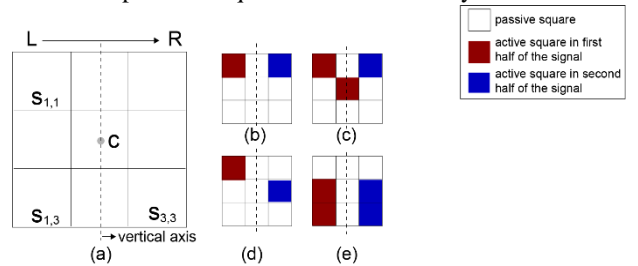

Fig. 11. (a) The area under hand is divided into nine equal squares, $s_{i,j}$ where $i$ and $j$ represent the column and row numbers. In the example shown above, the goal is to create a travelling vibrotactile flow from left to right. The active and passive squares shown in (b) and (e) are symmetric with respect to the vertical line, but (c) and (d) are not. Note that the line of symmetry is vertical (horizontal) for a horizontal (vertical) flow.

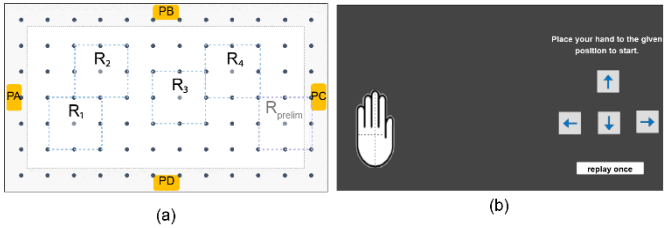

Fig. 12. (a) First, a preliminary experiment was conducted at the region $R_{prelim}$ and then the actual experiment was conducted at four different regions, $R_1$, $R_2$, $R_3$, and $R_4$. The size of each region was 12 by 12 cm. (b) Subjects were guided to align their hand position according to the hand image displayed on the screen. After the haptic stimulus was displayed, they were asked to determine the direction of vibrotactile flow by pressing one of the four arrow buttons on the screen.

The second experiment was conducted with eleven subjects (4 female, 7 male) having an average age of 29.6 years (SD: 6.0). Only two subjects were left-handed. The average hand width and length of the subjects were measured as 8.48 cm (SD: 0.52 cm) and 18.42 cm (SD: 6.24 cm), respectively. During the experiment, all subjects stood in front of HapTable and placed their hand on the five designated regions (one for preliminary and four for actual experiment), randomly distributed on the tabletop surface (Fig. 12a). They wore active noise-cancelling headphones playing white noise to block any auditory cues.

Experiment started with a preliminary session to help subjects familiarize with haptic stimuli and interface. It consisted of 40 trials (10 repetitions x 4 directions) performed in $R_{prelim}$ region (Fig. 12a). During this session, subjects could replay the stimulus as many times as they desired, and ask questions to the experimenter about the experimentational procedure. In the actual experiment, subjects completed 160 trials (4 regions x 4 directions x 10 repetitions). For each region, the flow directions were displayed in random order while the order was same for each subject. In regions $R_1$ and $R_2$, stimulus was applied to the left hand of subjects; whereas in other regions, $R_3$ and $R_4$, it was applied to their right hand. Subjects were allowed to replay the stimulus only once. At the end of each trial, they were asked to select the direction of vibrotactile flow by pressing one of the four arrow buttons, representing the flow directions of travelling up, down, left, and right, displayed on the screen (Fig. 12b).

The recognition rates of the subjects for all directions and regions are shown in Fig. 13. The average accuracy of the subjects for all directions was 90% (SD = 3.6%, Fig. 13a). Fig. 13b shows the regional recognition accuracy of all subjects (Mean = 90%, SD = 3.1%). A two-way repeated measures ANOVA was used to investigate the statistically significant effects of region and direction on recognition accuracy. Mauchly's test was applied to check the violation of sphericity assumption. If needed, the degrees of freedom were corrected using Greenhouse-Geisser correction. Finally, Bonferroni corrected post-hoc analysis was carried out to further investigate the statistical differences between the groups.

Two-way repeated measures ANOVA showed a statistically significant interaction between region and direction (p=0.006, $\eta^2_{partial}$ = 0.317). Analysis of simple main effects for direction revealed that the direction had a significant for region $R_4$ (p < 0.0005, $\eta^2_{partial}$ = 0.682), but not for other regions. Pairwise comparisons for this region showed that the difference in the recognition accuracy between left to right and remaining (down to up, up to down, and right to left) directions were statistically significant (p = 0.001, p = 0.005, and p= 0.002 respectively). Mode shape analysis revealed that during the first part of the stimulus displayed in this region for vibrotactile flow of left to right, there is a vibration on both sides of the hand, resulting in confusion about the direction (Fig. 14b).

## 5.3 Haptic Knob

In contrast to physical controls such as buttons, sliders, and knobs, virtual controls displayed on tabletops cannot be felt. As a result, task precision and performance drop [2]. Moreover, lack of haptic feedback requires continuous visual attention on the controller. For example, graphical knobs with visual detents (notches) are frequently used in tabletop displays to rotate a virtual object in the scene or select an item from a pull-down menu using a rotation gesture. Lack of haptic feedback makes it difficult for the user to precisely rotate the object or quickly select the item from the menu. Moreover, she/he cannot rest her/his fingers on the knob and focus on the virtual object or the menu.

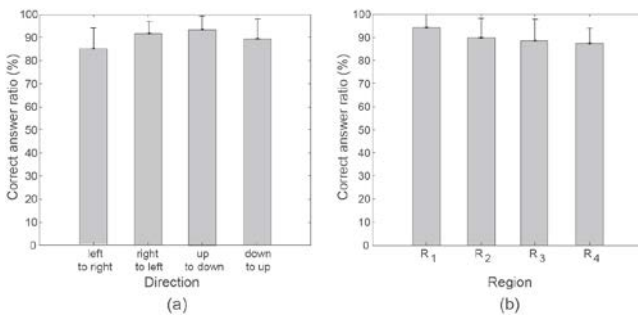

Fig. 13. Vibrotactile flow under hand: a) percentage of correct responses for each direction, b) percentage of correct responses for each region in the second experiment. The bars represent the mean values while deviations are the standard error of means.

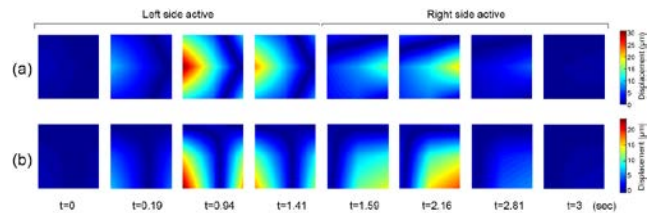

Fig. 14. Evolution of vibration maps for vibrotactile flow under hand: a) left-to-right direction at $R_3$ recognized with 95.5% accuracy (SD: 6.6%). The first and second parts of the input voltage signal successfully creates localized vibrations on the left and right sides of the hand sequentially, resulting in high accuracy in the subjects' perception of flow direction. b) left-to-right direction at $R_4$ recognized with 66.4% accuracy (SD: 17.7%). The first part of the signal vibrates the bottom left and right sides of the hand region at the same time, causing confusion in subject's perception about the direction of flow.

An alternative to a virtual control is a tangible control. In this approach, portable physical controls are directly placed on tabletop to augment visual interfaces with haptic feedback. These controls are detected with the help of touch sensing overlays or cameras. For example, Photo-Helix is a physical knob placed on a tabletop to interact with digital photo collection [27]. In this approach, one hand rotates the physical knob to control position on a helix-shaped calendar while the other hand inspects and modifies the digital photos. The translucent tangible knob in SLAP widgets can be used in various modes to interact with digital content depending on the application [28]. For example, the knob can be used in "jog wheel" mode to find and mark specific frames in a video or in "menu mode" to navigate through hierarchical menus. Weiss et al. used the knob in "jog mode" and conducted a user study with 10 participants. SLAP knob outperformed a virtual knob in terms of task completion time and accuracy. While tangible controls provide increased task performance, they have a fixed physical appearance unlike easily configurable virtual controls. Furthermore, they reduce the usable size of interaction area on tabletop surface and also restrict user movements.

## 5.4 Haptic Knob Experiment

We used real-time dynamic gesture recognition ability of our table and electrostatic actuation technique to display a haptic knob on our tabletop surface, a large size touch screen. In our experiments, subjects rotated the knob to navigate through a menu. We modulated the frictional forces between their fingers rotating the knob and the touch screen to investigate if haptic feedback improved their task completion time, precision, and subjective sense of accomplishing the task. To modulate the frictional forces, we applied voltage signal to the conductive layer of the touch screen in various forms as discussed below.

There were four sensory conditions (i.e. feedback types) in our study (Fig. 15):

1. *Virtual (V):* No artificial haptic feedback was displayed.
2. *Haptic Detent (HD):* A pulse signal was transmitted to the touch screen to generate a "notch" (detent) effect while the subjects crossed a sector during rotation (Fig. 15b). The purpose of haptic detent was to provide users with confirmation for the sector crossings, similar to a volume knob in a car.
3. *Haptic Detent and Constant Friction (HD+CF):* In addition to the pulse signals at sector crossings, a sinusoidal voltage signal with a constant amplitude (100 Vpp) and frequency (180 Hz, at which minimum electrovibration detection threshold for human finger was obtained for constant voltage by authors in [9]) was transmitted to the touch screen within the sector boundaries to display frictional haptic feedback during rotation for better control and precision (Fig. 15c).
4. *Haptic Detent and Velocity-based Friction (HD+VF):* The magnitude of the resistive frictional force was adjusted based on the subjects' angular velocity (the motivation for this type of haptic feedback stems from a rate-controlled joystick used in gaming applications, which simply displays more feedback force to faster movements). This was achieved by modulating the frequency of the input voltage applied to the screen between 60 and 180 Hz while keeping the amplitude constant at 100 Vpp.

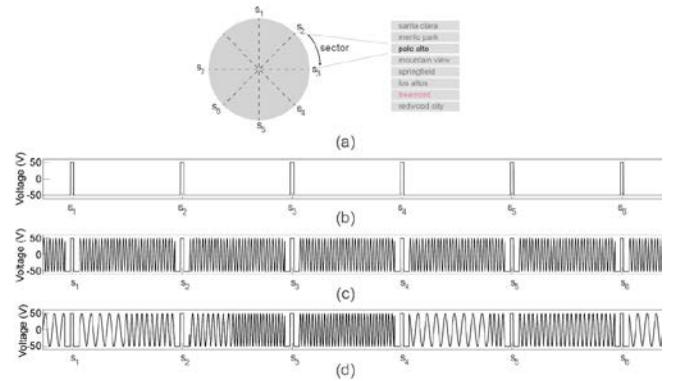

Fig. 15. (a) A knob with eight sectors where each sector is mapped to an item on the menu. In our experiments, subjects navigate on the menu under four different sensory conditions: 1) virtual (no artificial haptic feedback), 2) haptic detent at sector crossings (b), 3) haptic detent and constant friction (c), and 4) haptic detent and velocity-based friction (d).

Sixteen subjects (2 female, 14 male) with an average age of 29 years (SD: 5.2) participated in this experiment. Three subjects were left-handed, and none of the subjects had prior experience with electrostatic haptic feedback. They wore an antistatic wristband to their non-dominant hand, to connect their bodies directly to the ground, thus increasing the intensity of electrovibration. Subjects also put on active noise-cancelling headphones, playing white-noise, to block the environmental noise. The experiment took approximately sixty minutes to complete.

The experiment consisted of three consecutive sessions: (i) preliminary, (ii) testing, and (iii) subjective evaluation. Prior to the experimentation, all subjects were informed about the experimental procedure. The preliminary session helped subjects to get familiar with frictional haptic feedback displayed by electrovibration, rotation gesture, and the task itself (i.e. rotating the knob to navigate through a menu of items). A particular rotation gesture was chosen to provide comparable interaction experience across subjects. This gesture is performed with two fingers; thumb was the pivot point while index finger followed a circular arc.

During the experiment, subjects were asked to rotate the knob to navigate from a start city to a target city (marked with red color) on the menu consisting of randomized city names (Fig. 16a). As they navigated on the menu, a blue box highlighted the city that they were currently on. We investigated the effects of sector size, angular distance between start and target cities, and the senso-ry conditions on the task performance in terms of task completion time and task accuracy. Three different sector sizes (8, 16, 32 sectors) and three different angular distances between start and target cities (135, 270, 450 degrees) were used in the experiments. For a given angular distance, the number of cities on the menu and the frequency of haptic detents between start and target cities varied accordingly (since each sector always corresponded to one item on the menu). Subjects completed a total of 216 trials (4 sensory conditions x 3 sector sizes x 3 angular distances x 6 repetitions). The trials were displayed in random order while the order was same for all subjects.

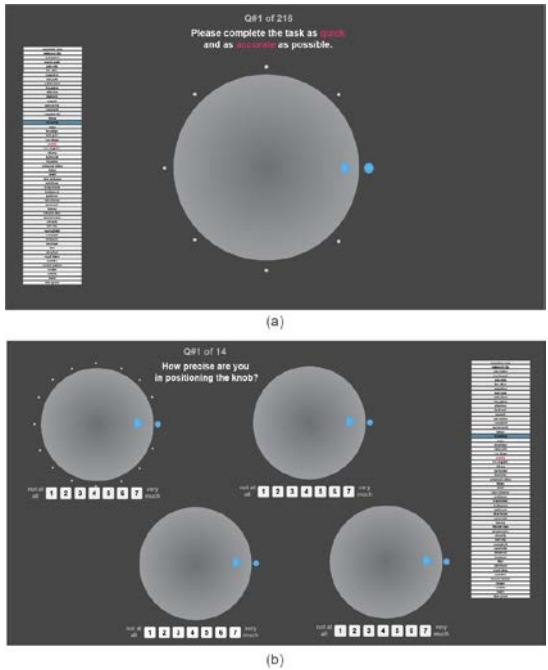

Fig. 16. (a) A knob with eight sectors used in the actual experiments, and the menu is shown on the left, (b) user interface for the subjective evaluation phase.

After the experiment, subjects were asked to fill a questionnaire, displayed digitally on the table surface, containing a total of fourteen questions (7 categories x 2 rephrased questions for each category). The questions aimed to measure their subjective experience under the four sensory conditions (V, HD, HD+CF, HD+VF). For each question, 4 knobs (one for each sensory condition) were displayed on the screen at the same time (Fig. 16b), allowing subjects to experience and compare the sensory conditions, and enter their experience for each condition using a 7-point Likert scale. As a reminder of the task performed in the actual experiment, we also provided the subjects with the menu (list of cities) in each question.

### 5.5 Results for the Haptic Knob Experiment

#### 5.4.1 Quantitative Results

To investigate the effects of sector size, angular distance between start and target cities, and the sensory conditions, we applied three-way repeated measures ANOVA on dependent variables of task completion time, overshoot rate, and recovery time. Mauchly's test of sphericity was first performed to check whether the differences between the levels of the within-subject factors have equal variance. If the sphericity assumption was violated, the degrees of freedom were corrected using Greenhouse-Geisser correction. Finally, Bonferroni corrected post-hoc analysis was carried out to investigate where the statistically significant differences between the levels of within-subject factors lie. The results for each quantitative metric can be summarized as follows:

1. *Task completion time* is the time it takes for a subject to navigate from start to target city in milliseconds. The results showed that there was no statistically significant three-way interaction (p=0.399). However, there was a significant two-way interaction between sector size and angular distance (p=0.013). Observations from simple main effects of these two factors showed that increasing either angular distance or number of sectors resulted in a statistically significant increase in task completion time. When angular distance was fixed, although the physical distance that subjects' fingers travel remained unchanged, subjects opted to slow down since they observed a greater number of cities had to be crossed between start and target cities. Type of sensory feedback did not influence task completion time.

2. *Overshoot rate* is the total number of times that a subject missed target city. Although there was no statistically significant three-way interaction, a statistically significant two-way interaction was observed between sector size and angular distance again (p=0.001). Increasing sector size or decreasing angular distance increased overshoot rate. Type of sensory feedback did not influence overshoot rate.

3. *Recovery time* is the time that it takes for a subject to reach target city after the first miss. Results showed that there was no significant interaction between any pairs of independent variables. However, sector size and angular distance had statistically significant main effect on the dependent variable (p=0.05). Subjects spent more time to recover the target when either sector size or angular distance was increased. Type of sensory feedback did not influence recovery time.

#### 5.4.2 Qualitative Results

The response of the subjects to the questions (subjective scores) are given in Fig. 17. To evaluate these responses, we used one-way repeated measures ANOVA. The results of ANOVA for each category in the questionnaire are summarized below:

- *Effectiveness:* Subjects reported that haptic knobs (HD, HD+CF, HD+VF) were significantly more effective than the virtual knob (V) (p=0.007).
- *Ease of use:* Although the subjective scores suggest that the haptic knob was easier to use compared to the virtual one, the difference between them were not statistically significant.

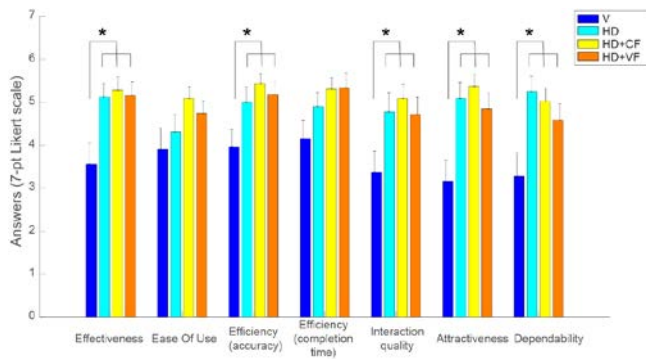

Fig. 17. Means and standard errors of the subjective measures for each sensory condition (* The mean difference is significant at p=0.05 level).

- *Efficiency (accuracy):* Subjects stated that they selected the target more accurately when haptic feedback was present (p=0.023). This outcome does not agree with the quantitative results for overshoot count and recovery time metrics.
- *Efficiency (time):* Although subjects personally stated that they completed the task fastest with knobs displaying HD+CF and HD+VF feedback, the subjective evaluation scores were not statistically significant. This outcome agrees with the quantitative results for task completion time.
- *Interaction quality:* Subjects perceived that the haptic knobs were more intuitive than the virtual knob during their interaction (p=0.026).
- *Attractiveness:* The results suggest that haptic knobs were more pleasant and attractive than the virtual one (p= 0.030).
- *Dependability:* Subjects felt greater confidence while completing the task with the haptic knobs, since they perceived them to be more dependable and supportive than the virtual one (p=0.013).

# 6 DISCUSSION

## 6.1. Vibrotactile Flow Experiments

In the first experiment, subjects were asked to put their index fingers of both hands on different locations on the table and then differentiate the direction of vibrotactile flow (either from the left index finger to the right one or vice versa). For a wide range of positional configurations, the subjects achieved perfect performance (100% correct identification) in this task that had a guess rate of fifty-percent. On the other hand, when they were asked to differentiate the direction of vibrotactile wave (up, down, left, right) travelling underneath their hand in the second experiment, the success rate dropped slightly. When rendering a vibrotactile flow underneath the subjects' hand, the rendering approach utilized for the first experiment had to be modified since the contact with the surface involved an area (hand) rather than a point (tip of an index finger). We divided the area under the user's hand into smaller regions and considered the amplitude of vibrations in those regions and their symmetry with respect to the horizontal and vertical axes dividing the area into two equal halves. Although this modification required extensive precomputations to identify the symmetric regions having a significant difference in vibration amplitude for all the excitation frequencies varied from 0 to 650 Hz, it was done only once. The results of the second experiment showed that the subjects could differentiate the direction of haptic flow with an average accuracy of 90% (SD = 3.6%) across the four directions. The slight drop in recognition accuracy compared to the first experiment is not suprising for several reasons. First of all, the sensitivity of index finger to vibrotactile stimulus is higher than palm [16, 17]. Second, the guess rate in the second experiment was 25% compared to that of 50% in the first experiment. Third, traveling vibrations can be better localized by index fingers of two separate hands rather than those traveling beneath one hand only. Finally, our current approach assumes a fixed-size area for human hand (within 12 x 12 cm square) and implements the haptic stimuli accordingly. Hence, a person with a hand smaller or larger than that assumed area could be in slight disadvantage in our current approach. The success rate for the directional flow can be further improved if the algorithm is auto-tuned with respect to the actual hand dimensions of the user. This can be accomplished using our gesture detection system that already extracts the hand contour and orientation.

## 6.2. Haptic Knob Experiment

The quantitative results of the haptic knob experiment showed that frictional haptic feedback did not really improve the task performance in terms of completion time, number of overshoots, and recovery time after the first overshoot, when the same metrics were compared to that of no artificial haptic feedback. This result may initially appear to be surprising since several earlier studies in other domains have showed that haptic feedback improves task performance. However, it is important to emphasize that there is a major difference between our study and the earlier ones. In those studies, subjects who performed the task under visual feedback condition did not receive any haptic feedback at all. In our study, although no artificial friction was displayed to the subjects under visual feedback condition, they still felt some amount of friction when their fingers performed the rotation gesture. It appears that the additional friction displayed by electrovibration did not help them much in executing the task faster and with less error. This outcome may be related to the type of rotation gesture used in our study. Two-finger rotation gesture already provides more control to a user due to constrained wrist motion and slower rotational speed. Voelker et al. reported that the task accuracy with virtual rotary knob controlled with two fingers is comparable to those of tangible knobs, while the task completion is longer [32]. However, we have chosen two-finger gesture because we observed in our initial experiments that the intensity of haptic feedback drops as the number of fingers increases.

Finally, the outcome of our haptic knob study may also be related to the way that frictional haptic feedback displayed to the subjects in our experiments. Although we have tried a wide range of alternatives, there are still many other options that can be explored in the future. For example, in HD condition, we utilized a pulse signal at sector crossings to imitate the feeling of detents. Alternatively, a detent can be, for example, rendered by leaving a gap between two subsequent pulses. A similar argument can be extended to the other haptic conditions. In our study, we utilized a sinusoidal voltage to display constant frictional haptic feedback under HD+CF condition. However, a sinusoidal voltage signal that is amplitude modulated to display more friction close to the sector boundaries (in order to slow down the rotational speed of user) could make improvements in task performance. As obvious from the short discussion above, there are several alternative choices for the design of haptic knob, which needs to be further explored in the future.

On the other hand, the subjective assessment following the knob experiments (via 14 questions in 7 different categories) showed that the subjects strongly preferred the haptic knobs over the virtual one in almost all categories (Fig. 17). The results are encouraging and suggest that adding haptic feedback to a virtual knob improves interaction quality, user experience, and also the confidence of user. For example, feeling the detents while rotating the knob does not perhaps help much in terms of task performance, but allows the user to receive a confirmation, which appears to improve her/his personal interaction experience and confidence.

## 7 CONCLUSION

This paper presents the design of a novel multimodal tabletop that effectively combines visual and haptic modalities to provide an interactive experience to a user. Users convey their intention of interaction via static and dynamic hand gestures [29], and HapTable recognizes these gestures in real time to display haptic feedback to the user accordingly. We demonstrated the haptic feedback capabilities of our table via two example applications along with detailed user studies; one for static and one for dynamic gestures. However, HapTable is not restricted to only these gestures and can be potentially used to display haptic feedback for a variety of other hand gestures, with applications in information and data visualization, games, entertainment, and education.

As an example for an interaction triggered by a static gesture, we displayed directional vibrotactile haptic feedback to the index fingers and hands of the subjects. Using the piezo patches attached to the edges of the tabletop surface, we successfully created an illusion of travelling vibrotactile flow beneath their fingers and hands in all four directions. We are not aware of any earlier study investigating vibrotactile flow on large-size touch screens such as ours. Using the methods presented in this paper, it is possible to display localized and directional vibrotactile haptic feedback on tabletop surfaces. Our future studies will investigate the potential applications of this technology in data visualization, education, and gaming. For example, in climate visualization, we imagine that a user can put their hand(s) on the tabletop surface to feel the direction of wind forces, virtually overlaid on some other graphical climate data. Considering the fact that climate data is complex and multi-dimensional, which overloads the visual channel, communicating some climate information, such as the wind forces, through haptic channel may alleviate the perceptual and cognitive load on the user, as suggested in [3]. Similarly, in an educational setting, a user could better appreciate granular materials such as sand, pebbles, beads, and seeds by shaking virtual cups containing them to feel the differences in their vibrations, rather than just observing their movements visually.

As an example for dynamic gestures, we haptically rendered a virtual knob on the table surface using the principles of electrostatic actuation. We investigated the potential benefits of frictional haptic feedback on task performance and user experience in selecting an item from a pull-down menu by rotating the knob. We are not aware of any earlier studies on electrostatic haptic rendering of a virtual knob on a touch surface. This required recognition of rotation gesture, tracking of individual finger positions, and displaying frictional forces to the user accordingly, all in real time. A knob is just one type of virtual control used in user interfaces and our future studies will investigate the haptic versions of the others such as slider, switch, button, and keyboard. For example, a haptic slider can be rendered by modulating the friction between the user's finger and the surface as in the case of haptic knob, while a key press can be simulated by localized vibrotactile effects using piezo patches. Once these controls are tested through user studies and designed as haptic widgets, they can be customized and integrated into various applications as a part of user interface.

In the applications mentioned above and the other potential ones, the challenge is to find the most effective mapping between the user hand gestures and the haptic effects. In fact, surface haptics is such a new area of research and even the more fundamental relations between the voltage signals applied to the actuators and our haptic sensing and perception are not well known yet. Without fully understanding those relations, developing an effective mapping between a gesture and haptic effect is highly challenging. For example, our recent work shows that electrovibration generated on a touch surface using a square voltage signal is perceived rougher than a sinusoidal one at low excitation frequencies [33]. For this reason, we preferred square pulses to render the detents of our haptic knobs over sinusoidal ones (Fig. 15) to make the sector crossings more detectable by the subjects. Finally, in our current study, only one type of haptic modality (either vibrotactile or electrovibration) was utilized to render haptic effects in each of the exemplar cases. However, the integration of two modalities on the same application may lead to richer haptic effects. For example, in our study, the detents of the knob could be displayed by vibrotactile feedback while frictional forces are conveyed to the user via electrovibration during the rotational movements.

**ACKNOWLEDGMENT**

The Scientific and Technological Research Council of Turkey (TUBITAK) supported this work under contract 114E0003 and student fellowship program BIDEB-2211.

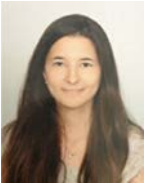
**Senem Ezgi Emgin** received her summa-cum laude B.S. degree in information systems engineering from Bogazici and SUNY Binghamton University in 2009, and MS degree in computer science from Binghamton University in 2011. She is currently a PhD student in computational sciences and engineering program at Koc University. Her research interests include surface haptics, user interfaces, graphics, visualization, and machine learning.

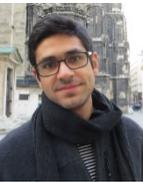
**Amirreza Aghakhani** received the B.S. degree in mechanical engineering from Sharif University, Tehran, Iran, in 2008. He is currently pursuing the Ph.D. degree in mechanical engineering at Koc University, Turkey. His current research interests include energy harvesting, piezoelectricity, impedance modeling, structural dynamics, vibrations, and haptics. Mr. Aghakhani is a member of the American Society of Mechanical Engineers (ASME), a student member of institute of electrical and electronics engineering (IEEE), and a member of the international society for optics and photonics (SPIE).

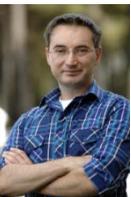
**Metin T. Sezgin** graduated summa-cum laude with honor's from Syracuse University in 1999. He received the MS and PhD degrees from the Massachusetts Institute of Technology in 2001 and 2006. He subsequently joined the University of Cambridge as a postdoctoral research associate, and held a visiting researcher position at Harvard University in 2010. He is currently an associate professor at Koc University, Istanbul and directs the Intelligent User Interfaces (IUI) Laboratory. His research interests include intelligent human computer interfaces and HCI applications of machine learning. He is particularly interested in applications of these technologies in building intelligent pen-based interfaces.

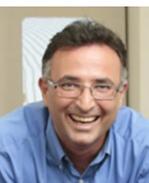
**Cagatay Basdogan** received the PhD degree in mechanical engineering from Southern Methodist University, in 1994. He is a faculty member in the mechanical engineering and computational sciences and engineering programs at Koc University, Istanbul, Turkey. He is also the director of the Robotics and Mechatronics Laboratory (RML), Koc University. Before joining Koc University, he worked at NASA-JPL/Caltech, MIT, and Northwestern University Research Park. His research interests include haptic interfaces, robotics, mechatronics, biomechanics, medical simulation, computer graphics, and multi-modal virtual environments. He is currently the associate editor in chief of the IEEE Transactions on Haptics and serves on the editorial boards of the IEEE Transactions on Mechatronics, Presence: Teleoperators and Virtual Environments, and Computer Animation and Virtual Worlds journals. He is a senior member of the IEEE.